\documentclass[aps,floatfix,eqsecnum,showpacs,preprint,tightenlines,nofootinbib]{revtex4-1}
\usepackage{epsfig}

\usepackage{amsmath}
\usepackage{dcolumn}% Align table columns on decimal point
\usepackage{bm}% bold math
\usepackage{hyperref}
\begin{document}
%\bibliographystyle{prsty} % Choose Phys. Rev. style for bibliography
% Use the \preprint command to place your local institutional report
% number in the upper righthand corner of the title page in preprint mode.
% Multiple \preprint commands are allowed.
% Use the 'preprintnumbers' class option to override journal defaults
% to display numbers if necessary
%\preprint{}
%\addbibresource{references.bib}
%Title of paper
\title{Correlations among neutron-proton and neutron-deuteron elastic scattering observables}
\author{Yu.~Volkotrub}
\affiliation{AGH University of Science and Technology, Faculty of Physics and Applied Computer Science, PL-30055 Krak\'ow, Poland}
\author{R. Skibi{\'n}ski}
\affiliation{M. Smoluchowski Institute of Physics, Jagiellonian University, PL-30348 Krak\'ow, Poland}
\author{J. Golak}
\affiliation{M. Smoluchowski Institute of Physics, Jagiellonian University, PL-30348 Krak\'ow, Poland}
\author{H. Wita{\l}a}
\affiliation{M. Smoluchowski Institute of Physics, Jagiellonian University, PL-30348 Krak\'ow, Poland}

\date{\today}

\begin{abstract}
We employ two models of the nucleon-nucleon force: the OPE-Gaussian as well as the chiral N$^{4}$LO and N$^{4}$LO$^{+}$ interactions 
with semilocal regularization in momentum space to study correlations 
among two-nucleon and three-nucleon elastic scattering observables. These models contain a number of free parameters whose 
values and covariance matrices are evaluated from a fit to the two-nucleon data.
Such detailed knowledge of parameters allows us to create, using various sets of statistically generated parameters, 
numerous versions of these potentials and next apply them to
two- and three-nucleon scattering to make predictions of various observables at the reaction energies up to 200 MeV.
This permits a systematic analysis of correlations among two-nucleon and three-nucleon observables, basing on 
a relatively big sample of predictions. We found that most observables in neutron-proton and neutron-deuteron systems are uncorrelated, 
but there are exceptions revealing strong correlations, which depend on 
the reaction energy and scattering angle.
This information may be useful for
precise fixing free parameters of two-nucleon and three-nucleon forces and for understanding dependencies 
and correlations between potential parameters and observables.
%Background, Purpose, Methods, Results, and Conclusions
\end{abstract}

% insert suggested PACS numbers in braces on next line
\pacs{21.45.-v, 25.10.+s, 13.75.Cs}
%25.10.+s Nuclear reactions involving few-nucleon systems
%25.40.Cm 	Elastic proton scattering
%21.45.-v 	Few-body systems
%13.75.Cs 	Nucleon-nucleon interactions
% insert suggested keywords - APS authors don't need to do this
%\keywords{}

%\maketitle must follow title, authors, abstract, \pacs, and \keywords
\maketitle

% body of paper here - Use proper section commands
% References should be done using the \cite, \ref, and \label commands

\section{\label{sec:level1}Introduction}
One of the challenges in nuclear physics is to describe ab initio, i.e. starting from reliable models of the
nucleon-nucleon (NN) and many-nucleon forces, properties of nuclear systems and reactions.
The current understanding of the nuclear forces 
is that they are residual interactions of the strong forces between quarks and gluons.
Nowadays we are not able to apply Quantum Chromodynamics (QCD) directly in the non-perturbative region and 
to describe processes at the nuclear scale starting from quarks and gluons and their interactions. 

Instead various effective models of nuclear interactions have been prepared. The Argonne v18 
(AV18)~\cite{AV18} and the CD-Bonn~\cite{CDBonn2001} models of NN interaction are important 
examples which provide a quite satisfactory description of numerous nuclear phenomena. The quality of two-nucleon (2N) 
data description can be measured by the $\chi^2$/datum value obtained from a comparison of theoretical 
predictions and experimental data.  
The AV18 and CD-Bonn forces achieved $\chi^2$/datum close to 1 but by using 40 and 45 free parameters, respectively. 
These parameters were fitted, via the phase shifts obtained by the Nijmegen group~\cite{Stoks}, to all 2N 
data available at that time. In both cases, only the central values of the parameters 
were determined and no information about their uncertainties, to the best of our knowledge, was published. 
A serious disadvantage of these models is the lack of a direct link to QCD. This issue does not occur in 
the models of nuclear forces derived within the $\chi$EFT approach~\cite{Epelbaum_review_2012} such as the recent potential
with the semi-local regularization in momentum space (SMS) proposed by the Bochum-Bonn group~\cite{Reinert2018}
or the forces derived by R. Machleidt \textit{et al.}~\cite{Entem2017},
M. Piarulli \textit{et al.}~\cite{Piarulli} and others~\cite{specialFrontiers}.

In the course of time, as the more and more advanced nuclear force models were derived, the accurate determination 
of their free parameters became more and more important. The crucial step was taken by R. Navarro P\'erez
and his collaborators from the Granada group. They revised the available 2N data and prepared a new 
database~\cite{Navarro2013}, rejecting experimental results for which 
the experimental uncertainties were unknown or poorly defined. They excluded also data sets inconsistent 
with other experimental results. This procedure led to consistency of the eventually accepted data sets. 
The Granada-2013 database is currently a standard collection of data used for fixing parameters of the NN forces. 

For some models of the NN interaction, derived in the 21st century, like the OPE-Gaussian 
force~\cite{Navarro2014} or the chiral SMS force~\cite{Reinert2018}, in addition to the central 
values of parameters also their covariance matrices were determined. This has opened up new possibilities
in few-nucleon studies. The propagation of the uncertainties of the potential parameters from 
a 2N system to many-body observables is one example. We have studied this issue
in~\cite{Skibinski2018, Volkotrub2020} and determined for the first time in a quantitative way the 
corresponding theoretical uncertainties for the elastic nucleon-deuteron (Nd) observables. 
Another interesting problem that can be investigated with the help of the covariance matrix of the potential parameters 
is the existence of correlations among various observables in two- and three-nucleon (3N) systems as well as 
between observables and specific potential parameters. In particular this can lead to 
establishing a set of observables that should (should not) be taken into account while 
fixing the free parameters of the three-nucleon force (3NF). The free parameters of the older 3NF models,
 like the Tucson-Melbourne~\cite{Coon1979, Coon1981, Coon2001} or the UrbanaIX~\cite{Pudliner1997} 
were determined from the $^{3}$H binding energy and the density of the nuclear matter. 
In the case of the chiral models, for which the 3NF occurs for the first time at N$^{2}$LO~\cite{oldBochum3NF,Epelbaum2005}, 
its two free parameters ($c_{D}$ and $c_{E}$) were initially fixed from the $^{3}$H binding energy and 
the neutron-deuteron ($nd$) doublet scattering length $^2$a$_{\rm{nd}}$~\cite{oldBochum3NF, Skibinski2011}. 
However, it is known that these two observables are strongly correlated and their linear dependence manifests itself 
in the so-called Phillips line~\cite{Phillips1968}. 
More recently, in Ref.~\cite{Maris2021}, beside the $^{3}$H binding energy, the differential cross 
section at $E_{\rm{lab}}= 70~\rm{MeV}$ around its minimum was used instead of $^2$a$_{\rm{nd}}$ 
for a 3NF consistent with the SMS regularization. The choice of the cross section was dictated by 
the significant 3NF effects observed in this angular region and by the existence of the very precise 
experimental data~\cite{Sekiguchi2002}. In Ref.~\cite{Epelbaum2019} authors discussed also the dependence 
of the $c_{D}$ and $c_{E}$ values on the choice of the energy and scattering angle ranges 
used during the fixing procedure, and found it small.

However, the question of possible correlation between 
the $^{3}$H binding energy and the scattering cross section is still open. The answer is important since 
obviously using the strongly correlated observables to fix free parameters can bias the results of 
such a determination method. Moreover, thirteen new free parameters, namely strengths of contact terms, for 3NF are expected at N$^4$LO~\cite{Pisa}. Fixing them 
will be a tremendous computational effort, and therefore the set of observables used for this purpose must be 
carefully selected. Specifically, to minimize uncertainties of fixed parameters, the selected observables 
should be uncorrelated. While the use of emulators proposed in~\cite{emulator1, emulator2} can reduce 
the amount of required CPU time, the accuracy of parameter values will still depend on a set of input observables.
Similarly, correlations discovered in the 2N system could impact the procedures used to fix free parameters 
of the NN force.  

In the past a study of correlations was not possible at a statistically significant level, 
due to the lack of a sufficiently large number of the realistic models of the nuclear potential and precise data. 
The situation has changed in recent years. Using the OPE-Gaussian or the chiral SMS forces allows us to prepare 
many sets of the potential parameters and thus, after a procedure described in the next section, obtain 
a number of predictions large enough to analyze correlations and to draw quantitative conclusions. 
Some attempts to study correlations in few-nucleon sectors are presented in Refs.~\cite{Perez2016},
~\cite{kirscher2010universal} and~\cite{kievsky2018correlations}. In Ref.~\cite{Perez2016} the authors study, 
using the Monte-Carlo bootstrap analysis as a method to randomize proton-proton and neutron-proton ($np$) scattering predictions, 
correlations between the ground states of the $^{2}$H, $^{3}$H and $^{4}$He binding energies, 
focusing mainly on the Tjon line~\cite{TJON1975217}, i.e. the correlation between $^{3}$H and $^{4}$He 
binding energies, but do not investigate the scattering observables. In Ref.~\cite{kirscher2010universal} 
the correlations between three- and four-nucleon observables have been investigated within the pionless 
Effective Field Theory with the Resonating Group Method. Because this approach can be applied only to 
processes at very low energies the authors focus on the study of bound-state properties and the $^{3}$H-neutron 
$S$-wave scattering length, finding the latter correlated with the $^{3}$H binding energy. 
Kievsky \textit{et al.}~\cite{kievsky2018correlations} studied correlations among the low-energy 
bound state observables in the two- and three-nucleon systems, 
extending 
this research to some features of the light nuclei and beyond up to nuclear matter and neutron star properties. 
Using a simple model of ``Leading-order Effective Field Theory inspired potential'' they found evidence of 
the connection between few- and many-nucleon observables. In addition, in Ref.~\cite{Escalante} correlations 
among partial wave dependent parameters in NN scattering, arising form the experimental data have been discussed 
in the context of uncertainty of models of the nuclear interaction. None of these works is focused 
on correlations in the context of three-nucleon observables. 

In the present paper we study correlations among observables in neutron-proton elastic scattering as well as among 
observables in neutron-deuteron elastic scattering 
focusing on the latter ones.
The paper is organized as follows: 
in Sec.~\ref{formalism} we list the essential steps of our formalism. In Secs.~\ref{in2N} and~\ref{in3N} 
we show results on correlations among various two- and three-nucleon observables, respectively. 
Finally, we summarize and conclude in Sec.~\ref{summary}. 

\section{Formalism}
\label{formalism}
The realization of our work can be split into the following phases:
\begin{enumerate}
\item Preparation of sets of the potential parameters

Having at our disposal the central values of the potential parameters and their covariance matrix,
we sample, in addition to set S$_0$ of central values of potential parameters,  
50 sets S$_i, i=1,2,\dots,50$ of the potential parameters from the multivariate normal distribution. 

\item Computing observables for each set of potential parameters

All results in this work have been obtained in momentum space by resorting to standard partial wave decomposition (PWD).
We neglect the Coulomb force and the 3N interaction. 
For each set $S_{i}$ $(i = 0,1, \ldots, 50)$ defining a NN interaction $V$, we compute the deuteron wave function 
by solving the Schr{\"o}dinger equation. 
Next, for the same $V$ we solve the Lippmann-Schwinger equation, $t = V + V\tilde{G}_{0}t$ ($\tilde{G}_{0}$ is the free
2N propagator), to obtain the 2N $t$-matrix, from which 2N scattering observables are
obtained. For 3N system we solve the Faddeev equation and construct the transition amplitudes from which the 3N scattering 
observables are computed, see Chap.~\ref{in3N}. 
For more details on these calculations we refer the reader to Refs.~\cite{Glockle99109, GLOCKLE1996107} 
and references therein. As a result, an angular dependence of various $np$ and $nd$ elastic scattering observables is 
obtained for each set of parameters $S_{i}$. 

The resulted predictions can be used to study:
\begin{enumerate} %[label=(\alph*)] % (a), (b), (c), ...
\item for a given observable $X$ at an energy $E$ and at a scattering angle $\theta$, the empirical 
probability density function of the observable $X$($E$, $\theta$) resulting when various sets 
$S_i$, ($i = 0, \ldots, 50$) are used;
\item for a given observable $X$, both the angular and energy dependencies of predictions based 
on various sets $S_{i}$.

\hspace{-1cm} This, in turn, allows us to analyze correlations among all observables.   

\item We calculate a sample estimator of the correlation coefficient between two chosen observables $X$ and $Y$, using the standard formula
\begin{equation}
r(X,Y) = \frac{\sum\limits_{i=1}^{n}\left(x_{i} - \bar{X}\right)\left(y_{i} - \bar{Y}\right)}{\sqrt{\sum\limits_{i=1}^{n}\left(x_{i} - \bar{X}\right)^{2}\sum\limits_{i=1}\left(y_{i} - \bar{Y}\right)^{2}}}\;,
\label{correlation}
\end{equation}
where the
index $``i"$ runs over the sets of $n = 51$ versions of potentials, $\bar{X}$ and $\bar{Y}$ are means: $\bar{X} = \frac1n \sum\limits_{i=1}^{n}x_{i}$ and $\bar{Y} = \frac1n \sum\limits_{i=1}^{n}y_{i}$, respectively.

\end{enumerate}

\end{enumerate}

The interpretation of the correlation coefficient is to some extent arbitrary.
The correlation coefficient takes values from -1 to 1 and we define 
the case $\vert r \vert \le 0.3$ as no correlation, while $0.3 < \vert r \vert \le 0.5$, 
$0.5 < \vert r \vert \le 0.7$, and $0.7 < \vert r \vert \le 1$  mean weak, moderate, and strong correlation, respectively. 
Specifically, $\vert r(X,Y) \vert =1$ means linear dependence between the two observables $X$ and $Y$.
If the correlation coefficient equals 0, then there is no linear dependence between the observables, 
however, they can still be nonlinearly dependent.

In the determination of the correlation coefficient for a given pair of observables, we have 
in practice only one 50-element sample at our disposal. Estimation of uncertainty of 
computed sample correlation coefficients is not straightforward. To find this uncertainty we apply 
two methods. First we use the well-known Fisher transformation~\cite{Fisher1915, Fisher1921}. To construct 
a confidence interval by Fisher's method, it is assumed that $X$ and $Y$ have a bivariate normal 
distribution. This assumption is only approximately satisfied by observables, see Ref.~\cite{Skibinski2018} for the corresponding discussion 
on 3N observables obtained with the OPE-Gaussian potential. 
The variable $Z = \frac{1}{2}\rm{ln}\left(\frac{1 - r(X,Y)}{1 + r(X,Y)}\right)$, which has 
the standard normal distribution is used to construct a confidence interval at a given confidence 
level $\gamma$. After inverse transformation of the confidence limits one gets the confidence interval 
for $r(X,Y)$. Applying this method to various pairs $(X,Y)$, we found that the obtained half confidence 
interval at $\gamma = 0.9$ is usually $\approx 0.05$ for $r \approx 0.9$ and $\approx 0.25$ for $r \approx 0.25$. 

Secondly, we use the bootstrap resampling method~\cite{eforn1979bootstrap,ikpotokin2013correlation,
bishara2017confidence} to estimate uncertainties of $r(X,Y)$. The advantage of the bootstrap method lies
in working directly with sample elements without any assumption about the normality of the $X$ and $Y$ distributions. 
Resampling (up to 1000 $(X, Y)$-elements) allows us to find the properties (e.g. distribution or 
standard derivation) of the bootstrap estimator for the correlation coefficient and thus estimate the uncertainty 
of originally received $r(X, Y)$, again as a half of bootstrap confidence interval. Typically, the half of 
the bootstrap confidence interval at $\gamma = 0.9$ is $\approx 0.05$ for $r \approx 0.9$ ($\approx 0.23$ 
for $r \approx 0.25$). These values are similar to uncertainties resulting from the Fisher method.

This shows that the uncertainty of the correlation coefficients, especially for small $r(X,Y)$,
is relatively high. Thus in the following we will restrict ourselves to a more qualitative discussion of correlation coefficients.
Since we are interested in finding whether given observables are or are not correlated,
such a qualitative result is sufficient.

\section{Correlations among neutron-proton elastic scattering observables}
\label{in2N}

Elastic scattering of two spin 1/2 particles offers much more diverse measurements than only 
of the differential cross section, $d\sigma/d\Omega$, since beside the unpolarized cross section various
spin observables are available.
In this chapter we present a few observables: the asymmetry $A^{\prime}$, the polarization $P$, 
the depolarization $R$, and the spin transfer coefficient $D$~\cite{Glockle99109}. We determine their correlation coefficients 
at two incident neutron en\-er\-gies $E_{\rm{lab}} = 10$, and 135 MeV in the range of the scattering angle 
$\theta_{c.m.} \in [12.5^{\circ}, 167.5^{\circ}]$~\footnote{This interval was chosen to avoid divergences 
occurring due to division by a very small value or by zero for $\theta_{c.m.} = 0^{\circ}$ or 
$\theta_{c.m.} = 180^{\circ}$ where the variance of observable tends to zero} using 
the chiral N$^{4}$LO and N$^{4}$LO$^{+}$ SMS NN potentials 
with the cutoff parameter $\Lambda = 450~\rm{MeV}$
as well as the OPE-Gaussian NN force.

\begin{figure}[ht]
\centering
\includegraphics[scale=0.7,clip=true]{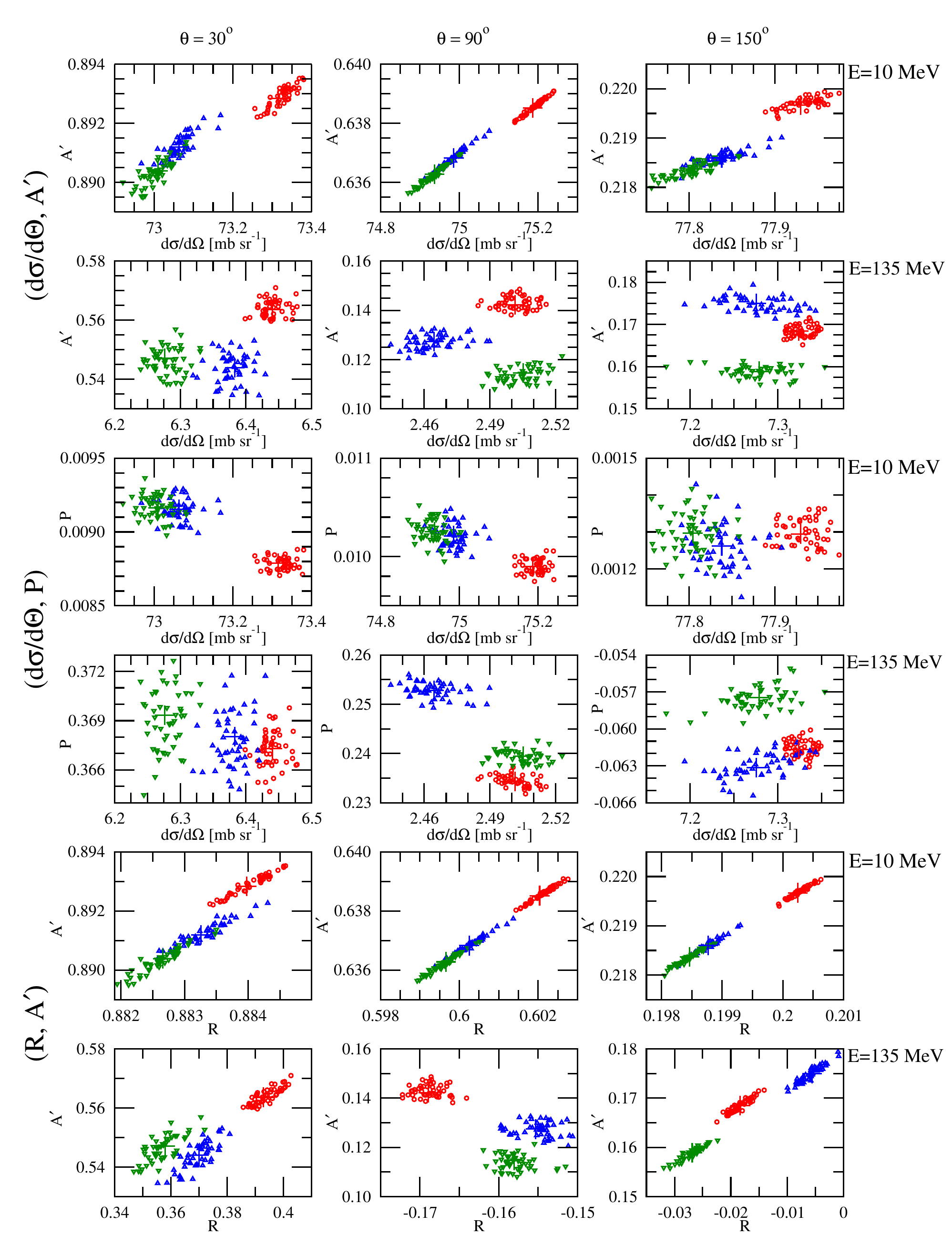}
\caption{Scatter plots for the chosen $np$ scattering observables at the c.m. scattering angle 
$\theta_{c.m.} = 30^{\circ}$ (left), $\theta_{c.m.} = 90^{\circ}$ (middle), and $\theta_{c.m.} = 150^{\circ}$ (right) 
and at the incoming neutron laboratory energy $E_{\mathrm{lab}} = 10~\mathrm{MeV}$ (odd rows) 
and $E_{\mathrm{lab}} = 135~\mathrm{MeV}$ (even rows). 
Two top rows are for the differential cross section $d\sigma/d\Omega$ and the asymmetry $A^{\prime}$, two middle rows 
are for the $np$ differential cross section $d\sigma/d\Omega$ and the polarization $P$ and two bottom rows are for 
the $np$ depolarization $R$ and the asymmetry $A^{\prime}$. 
The blue up-pointing triangles, green down-pointing triangles and red circles show predictions of 50 sets of potential parameters of the 
chiral N$^{4}$LO ($\Lambda = 450~\mathrm{MeV}$), chiral N$^{4}$LO$^+$ ($\Lambda = 450~\mathrm{MeV}$) SMS force 
and the OPE-Gaussian potential, respectively. The single pluses show  
predictions obtained with the central values of these potentials.}
\label{fig1:scatt_2N}       % Give a unique label
\end{figure}

In Fig.~\ref{fig1:scatt_2N} we demonstrate, in the form of scatter plots, 
predictions for the chosen 2N scattering observables based on these three models of the NN potential,
using, for each force, 51 sets of potential parameters. 
The top row in Fig.~\ref{fig1:scatt_2N} 
visualizes a strong positive correlation between $d\sigma/d\Omega$ and $A^{\prime}$ 
at $E_{\mathrm{lab}} = 10~\mathrm{MeV}$ at three scattering angles 
$\theta_{c.m.} = 30^{\circ}$, $90^{\circ}$, and $150^{\circ}$. The correlation coefficients for the chiral N$^{4}$LO SMS force
are $r(\theta_{c.m.}=30^{\circ})=0.81$, $r(\theta_{c.m.}=90^{\circ})=0.99$, $r (\theta_{c.m.}=150^{\circ})=0.85$ and
$(r(\theta_{c.m.}=30^{\circ})=0.87$, $r(\theta_{c.m.}=90^{\circ})=0.99$,
$r(\theta_{c.m.}=150^{\circ})=0.74$ for the OPE-Gaussian potential, so we conclude that the $d\sigma/d\Omega$ and $A^{\prime}$
are rather strongly correlated at this energy. For $E_{\mathrm{lab}} = 135~\mathrm{MeV}$ 
and all three scattering angles, the scatter plots in the second row of Fig.~\ref{fig1:scatt_2N},
show, that the strong correlation observed at $E_{\mathrm{lab}} = 10~\mathrm{MeV}$ disappears at higher energy for all the potentials,
leaving only  
weak correlation between $(d\sigma/d\Omega$ and $A^{\prime})$.

An analysis of two medium rows in Fig.~\ref{fig1:scatt_2N} leads to conclusion that  
$d\sigma/d\Omega$ is weakly correlated with the polarization $P$, independently of the 
NN potential used, the scattering energy, and the scattering angle (in fact, that is true 
for the entire $\theta_{c.m.})$ interval. Yet another pattern of correlations occurs 
for the $(R, A^{\prime})$ pair, see two bottom rows of Fig.~\ref{fig1:scatt_2N}. While at 
$E_{\mathrm{lab}} = 10~\mathrm{MeV}$ there is strong positive correlation for all the three scattering angles, 
at $E_{\mathrm{lab}} = 135~\mathrm{MeV}$ correlation depends on the scattering angle, changing from 
strong positive to almost negligible one. Once again we see that the three models of the NN interaction yield similar predictions for the correlation coefficients.
  
\begin{figure}[ht]
\centering
\includegraphics[scale=0.7,clip=true]{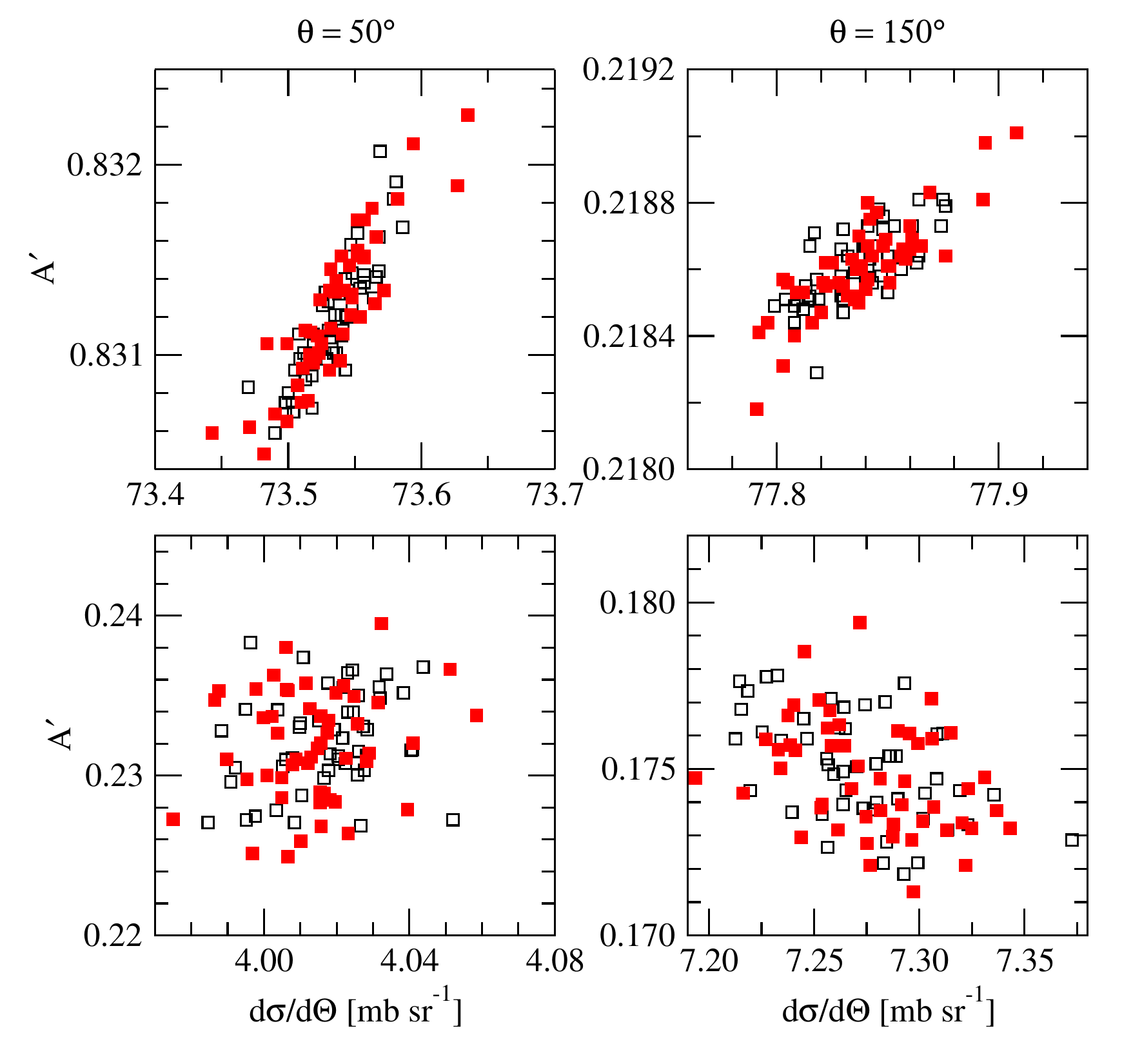}
\caption{
Scatter plots for $d\sigma/d\Omega$ and $A^{\prime}$ in $np$ scattering at the c.m. scattering angle
$\theta_{c.m.} = 50^{\circ}$ (left) and $\theta_{c.m.} = 150^{\circ}$ (right) and at the incoming neutron laboratory energy
$E_{\mathrm{lab}} = 10$~MeV (top) and $E_{\mathrm{lab}} = 135$~MeV (bottom) based on two 50-elements sets of potential parameters, represented by 
the open black squares and the full red squares.
The chiral SMS potential at N$^4$LO with $\Lambda=450$~MeV was used.
}
\label{fig2:samplesize}       % Give a unique label
\end{figure}

In the case of 2N system we are able to check
how our estimation of the correlation coefficients $r(X,Y)$ depends on the sample size
and find this dependence very weak.
For example a correlation coefficient $r_{50}(X,Y)$ obtained
from the 50-element sample calculated with the  N$^4$LO SMS force and
the corresponding correlation coefficient $r_{100}(X,Y)$ derived
from the 100-element sample have similar values.
In particular, at $E_{\mathrm{lab}} = 10$~MeV and $\theta_{c.m.} = 50^{\circ}$
the correlation coefficient $r_{50}(d\sigma/d\Omega,R)=0.95$ while $r_{100}(d\sigma/d\Omega,R)=0.93$. At $\theta_{c.m.} = 150^{\circ}$
$r_{50}(d\sigma/d\Omega,R)=0.82$ and $r_{100}(d\sigma/d\Omega,R)=0.80$. Similarly, for $r(d\sigma/d\Omega,A^{\prime})$ we get
$r_{50}(d\sigma/d\Omega,A^{\prime})=0.88$ and $r_{100}(d\sigma/d\Omega,A^{\prime})=0.87$ at $\theta_{c.m.} = 50^{\circ}$ and
$r_{50}(d\sigma/d\Omega,A^{\prime})=0.83$ and $r_{100}(d\sigma/d\Omega,A^{\prime})=0.78$ at $\theta_{c.m.} = 150^{\circ}$.
At $E_{\mathrm{lab}} = 135$~MeV the correlation coefficients are much smaller:
at $\theta_{c.m.} = 50^{\circ}$ $r_{50}(d\sigma/d\Omega,R)=0.29$ and $r_{100}(d\sigma/d\Omega,R)=0.37$,  
at $\theta_{c.m.} = 150^{\circ}$ $r_{50}(d\sigma/d\Omega,R)=-0.40$ while $r_{100}(d\sigma/d\Omega,R)=-0.44$.
For $r(d\sigma/d\Omega,A^{\prime})$ we have 
$r_{50}(d\sigma/d\Omega,A^{\prime})=0.24$ and $r_{100}(d\sigma/d\Omega,A^{\prime})=0.27$ at $\theta_{c.m.} = 50^{\circ}$ and
$r_{50}(d\sigma/d\Omega,A^{\prime})=-0.35$ and $r_{100}(d\sigma/d\Omega,A^{\prime})=-0.41$ at $\theta_{c.m.} = 150^{\circ}$.
While a small (usually about 5\%) difference between $r_{50}$ and $r_{100}$ is seen,
the qualitative conclusion about correlation coefficients 
remains unchanged with the increasing sample size. 
The independence from the sample used is also exemplified 
%This is also illustrated 
in Fig.~\ref{fig2:samplesize} for $d\sigma/d\Omega$ and $A^{\prime}$,
where two samples of size 50 yield a similar distribution of predictions.

\begin{figure}[ht!]
\centering
\includegraphics[scale=0.7,clip=true]{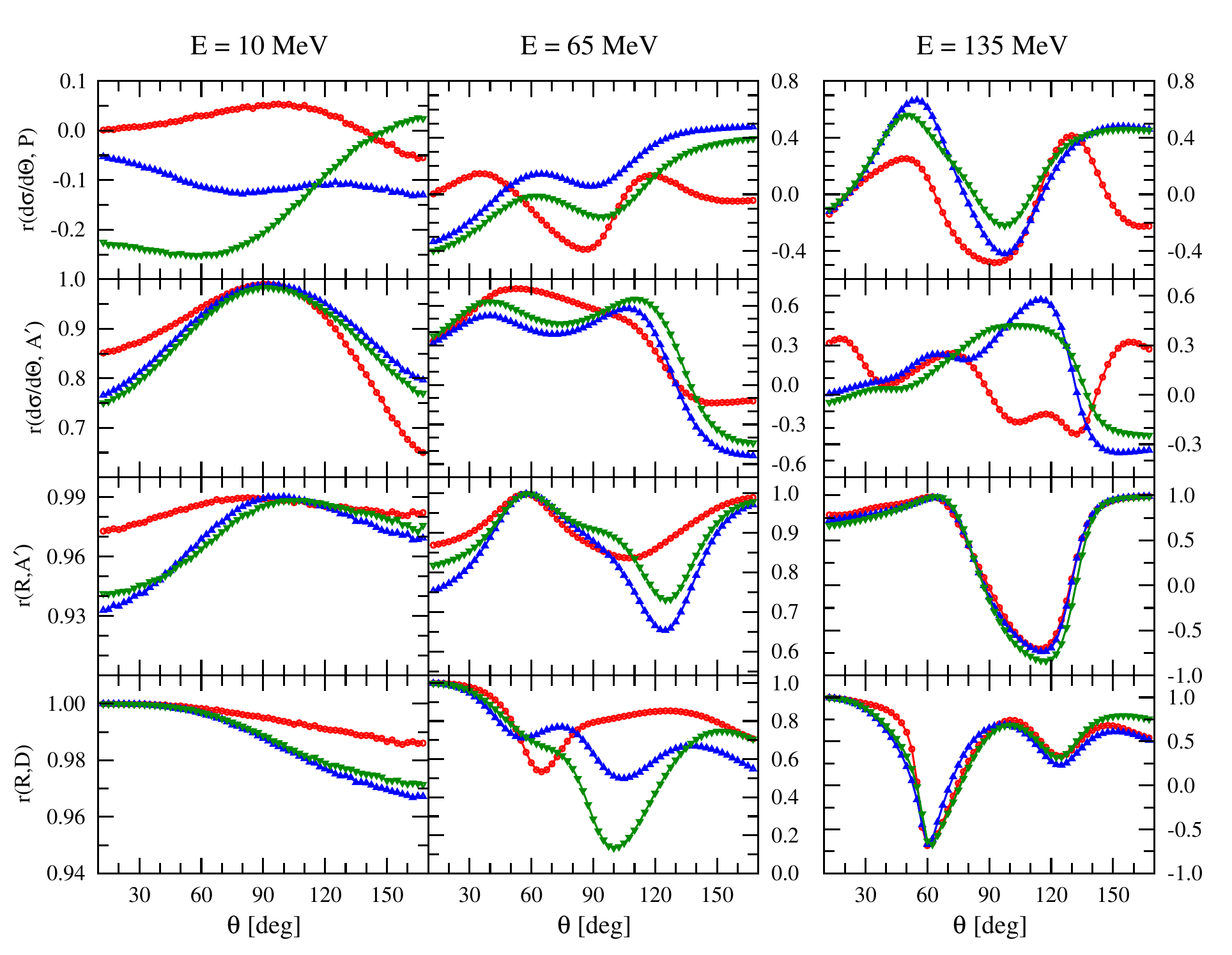}
\caption{
The angular dependence of correlation coefficients $r(d\sigma/d\Omega,P)$ (top), 
the $r(d\sigma/d\Omega, A^{\prime})$ (the second row), $r(R, A^{\prime})$ (the third row), and $r(R,D)$ (bottom) in $np$ scattering 
at the incoming neutron laboratory energy $E_{\rm{lab}} = 10~\rm{MeV}$ (left), $E_{\rm{lab}} = 65~\rm{MeV}$ (middle), and 
$E_{\rm{lab}} = 135~\rm{MeV}$ (right). 
The red curve with circles, the blue curve with up-pointing triangles, and the green curve with down-pointing triangles represent predictions of the OPE-Gaussian,
the SMS N$^4$LO, and the SMS N$^4$LO$^+$ interactions, respectively.
}
\label{fig2:angular2N}
\end{figure}

Focusing on the correlation coefficients we are able now to study their dependence on the scattering angle and the reaction energy
in a more compact form than shown in Fig.~\ref{fig1:scatt_2N}. 
Figure~\ref{fig2:angular2N} shows the angular dependence of four correlation coefficients 
for three energies of the incoming nucleon: $E_{\mathrm{lab}} = 10, 65,$ and 135 MeV.
Note the range of the Y-axis in Fig.~\ref{fig2:angular2N} does not cover the full possible range of the correlation 
coefficient $r \in [-1,1]$ otherwise many curves would be indistinguishable.
The top row, for $r(d\sigma/d\Omega,P)$, shows that at the lowest energy these observables remain uncorrelated for all
presented scattering angles, despite differences among predictions based on the considered interactions.
At the two higher energies the pattern of angular dependence is more complex and beside ranges of very weak or medium correlation
for both chiral models $d\sigma/d\Omega$ and $P$ are strongly correlated around $\theta = 50^{\circ}$ at $E_{\mathrm{lab}} =135$~MeV. 
As seen from the second row of Fig.~\ref{fig2:angular2N}, the $np$ differential cross section $d\sigma/d\Omega$ is, 
in general, strongly correlated with the asymmetry $A^{\prime}$ over a wide range of the scattering angle 
at the incoming neutron laboratory energy $E_{\mathrm{lab}} = 10$~MeV.
A magnitude of the $r(d\sigma/d\Omega,A^{\prime})$ exceeds at this energy $0.8$ in a wide range of the scattering angle.
For the two higher energies ($E_{\mathrm{lab}} = 65$ and 135 MeV) the correlation decreases, but still at 
some regions remains moderate. 
Figure~\ref{fig2:angular2N} shows also the angular dependence of the correlation coefficient 
for the $(R, A^{\prime})$ pair, which is strongly correlated for all interactions
at $E_{\mathrm{lab}} = 10$. At $E_{\mathrm{lab}} = 65~\mathrm{MeV}$, the magnitude of the correlation 
coefficient still has high values, but in the interval $\theta_{c.m.} \in (100^{\circ}, 150^{\circ})$ 
reaches its minimum of $r\approx 0.65$ for the chiral N$^{4}$LO SMS potential ($r\approx 0.75$ for N$^{4}$LO$^{+}$). 
At $E_{\mathrm{lab}} = 135~\mathrm{MeV}$, strong correlation is observed at 
$\theta_{c.m.} \in (50^{\circ}, 70^{\circ})$ and only negative moderate 
correlation occurs at $\theta_{c.m.} \in (110^{\circ}, 130^{\circ})$. 
Finally, in the last row of Fig.~\ref{fig2:angular2N} we display $r(R,D)$, which is close to 1 for the lowest energy,
but changes with the increasing energy, ending at $E_{\mathrm{lab}} = 135~\mathrm{MeV}$ 
in moderate positive correlation with an exception around $\theta_{c.m.}=60^{\circ}$, where 
a moderate negative correlation is observed.
 
Combining all the given above observations for the correlation coefficients among 2N observables 
and not shown here results for other pairs of 2N 
observables we conclude:

\begin{enumerate}
\item we observe a complex dependence of the correlation coefficients on the scattering angle and this complexity grows with the reaction energy.
While at small energies there are many correlated observables, this picture changes for the higher energies, where more and more observables
become uncorrelated. We suppose this effect is related to a bigger number of partial waves (and thus also potential parameters) 
contributing at higher energies.
Due to this complex pattern and relatively big uncertainties of the correlation coefficients, 
especially for small correlation coefficients, we restrict ourselves to only
qualitative conclusions.

\item correlation coefficients predicted from various models of the NN interaction, that is from the chiral potentials at various orders and the more 
phenomenological OPE-Gaussian force, are in qualitative agreement.
The observed differences are well inside the uncertainty of the obtained correlation coefficients. Thus, in practice it is possible 
to conclude about the correlation strength using a sample of parameters from only one model of interaction.

\item the polarization $P$ is weakly correlated with the other 2N scattering observables. This is true for all NN potentials 
(in the case of the chiral SMS forces starting from N$^{3}$LO), the scattering energies and scattering angles. 

\item the differential cross section $d\sigma/d\Omega$ is strongly correlated with all 2N scattering observables, except for $P$, at energies 
up to $E_{\mathrm{lab}} = 30~\mathrm{MeV}$, in specific intervals of $\theta_{c.m.}$. 

\item a sample of 50 sets of potential parameters is sufficient to study correlations between 2N observables. Such a study was impossible 
with availability of only few models of the NN potential for which only central values of parameters were known.

\end{enumerate}

\section{Correlations among neutron-deuteron elastic scattering observables}
\label{in3N}

We investigate the 3N system within the Faddeev formalism~\cite{GLOCKLE1996107,Glockle99109,Witala}. 
In the following we will just briefly describe the key equations used in this work.

Since we also studied correlations of scattering observables with the $^3$H binding energy, beside the scattering states we also
computed the $^3$H bound state for all $S_i$ sets of the potential parameters. 
The Faddeev-component $\vert \psi_{1} \rangle$ of the 3N bound state $\vert \Psi \rangle$ fulfills~\cite{Nogga1997}
\begin{equation}
\vert \psi_{1} \rangle = G_{0} t P \vert \psi_{1} \rangle \;,
\label{Eqbound}
\end{equation}
where the two-nucleon $t$-operator, defined in the previous section, acts now in the 3N space. 
$G_{0}$ is the free 3N propagator and $P$ is a permutation operator
built from transpositions $P_{ij}$: $P= P_{12} P_{23} + P_{13} P_{23}$. 
The complete 3N bound state $\vert \Psi \rangle = (1+P) \vert \psi_{1} \rangle $. 

The transition amplitude $U$ for the Nd elastic scattering is calculated prior to computing 3N scattering observables. 
Its matrix elements between the initial $\vert \phi \rangle$ and final $\vert \phi^{\prime} \rangle$ Nd states, neglecting 3NF, 
are given by~\cite{GLOCKLE1996107}
\begin{equation}
\begin{split}
\left<\phi^{\prime}|U|\phi\right> = \left<\phi^{\prime}|PG^{-1}_{0}|\phi\right> + \left<\phi^{\prime}|PT|\phi\right>\;.
\end{split}
\label{FaddeevElastic}
\end{equation}

The Faddeev equation for the auxiliary state $T\vert \phi \rangle$ with nucleons interacting only via a NN interaction $V$ entering a $t$-matrix 
expresses as~\cite{GLOCKLE1996107}
\begin{equation}
\begin{split}
T\vert \phi \rangle &= t P \vert \phi \rangle + tP G_{0}T\vert \phi \rangle\;.
\label{Faddeev2}
\end{split}
\end{equation}
In the above equations the initial and final Nd states 
are products of the deuteron state $\vert \varphi_{d} \rangle$ 
and a relative momentum eigenstate of the free nucleon, $\vert \vec{q}_{0} \rangle$, 
$\vert \phi \rangle = \vert \varphi_{d} m_{d} \rangle \vert \vec{q}_{0} m_{N} \rangle$ 
with the corresponding spin quantum numbers $m_{d}$ and $m_{N}$, respectively. 

Applying partial wave decomposition to operators and states in Eqs.~(\ref{Eqbound}) and~(\ref{Faddeev2}) we solve them numerically,
generating for Eq.~(\ref{Faddeev2}) its Neumann series and summing it up using the Pad\`e method~\cite{Glockle99109, GLOCKLE1996107}. 
The 3N partial wave basis comprises all states with the two-body subsystem total angular momentum 
$j$ $\leq$ 5 and the total 3N angular momentum $J$ $\leq$ $\frac{25}{2}$. This guarantees convergence of predictions 
with respect to these total angular momenta. The total number of the 3N states for given $J^\pi$ (total 3N angular momentum
with parity $\pi$ equal +1 or -1) amounts up to 142. 
We use grids of 32 $p$ points in the range 0-40 $\mathrm{fm}^{-1}$ and 37 $q$ points in the range 0-25 $\mathrm{fm}^{-1}$ both 
for the chiral SMS force and the OPE-Gaussian potential. From Eq.~(\ref{FaddeevElastic}) one finds matrix elements of the elastic 
transition amplitude $U$, from which a large set of observables can be computed~\cite{GLOCKLE1996107}. 
In total, there are 55 different observables for elastic scattering comprising the unpolarized cross section $d\sigma/d\Omega$,
nucleon and deuteron analyzing powers, spin correlation coefficients and polarization transfer coefficients from nucleon/deuteron to 
nucleon/deuteron; for more details 
see Ref.~\cite{GLOCKLE1996107}. It gives $\frac12 \times 55 \times 54 = 1485$ pairs of observables, so in the following we 
choose and describe only a few examples.

We present results for 3N observables in the same way as in the 2N case, i.e. first we discuss scatter plots and next move to 
the angular dependence of the correlation coefficients.

\begin{figure}[ht]
\centering
\includegraphics[scale=0.7,clip=true]{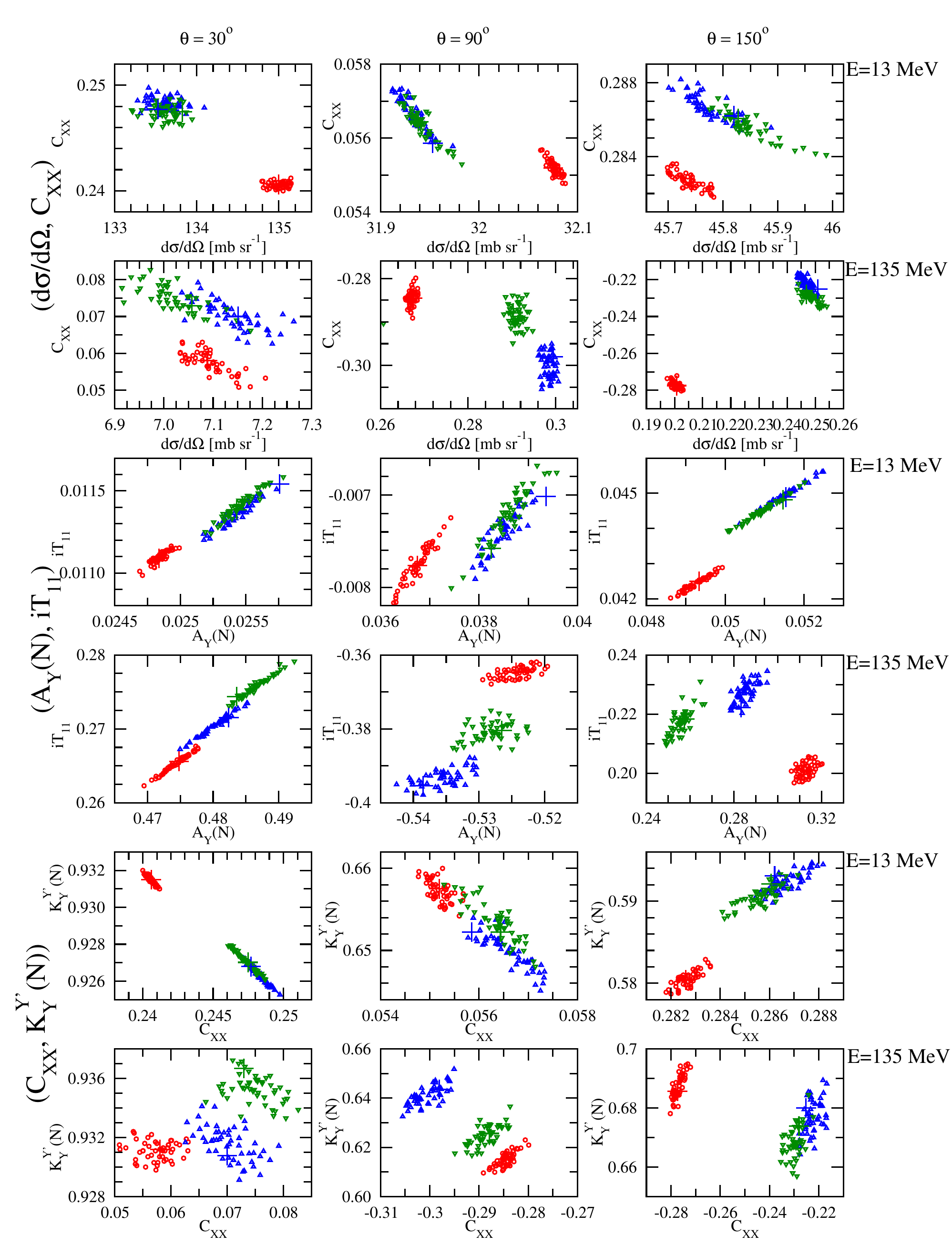}
\caption{A scatter plot between $(d\sigma/d\Omega, C_{xx})$, %$(C\s{YY},C\s{ZZ})$, 
$(A_{y}(n), \rm{iT}_{11})$, and $(C_{xx}, K^{y^{\prime}}_{y}(n))$ 
in the elastic $nd$ scattering at the c.m. scattering angle $\theta_{c.m.} = 30^{\circ}$ (left), $\theta_{c.m.} = 90^{\circ}$ (middle), 
and $\theta_{c.m.} = 150^{\circ}$ (right) and at the incoming neutron laboratory energy $E_{\mathrm{lab}} = 13~\mathrm{MeV}$ (odd rows) 
and $E_{\mathrm{lab}} = 135~\mathrm{MeV}$ (even rows). The red circles, blue up-pointing triangles and the green down-pointing 
	triangles represent results 
obtained with the OPE-Gaussian force, the N$^{4}$LO SMS and the N$^{4}$LO$^{+}$ SMS potentials, respectively. Pluses in the same colors shows 
predictions obtained with central values of the parameters.}
\label{fig4:scatt_3N}
\end{figure}

The scatter plots for the differential cross section $d\sigma/d\Omega$ and the spin correlation coefficient $C_{XX}$ 
are shown in two upper rows of Fig.~\ref{fig4:scatt_3N}
for three scattering angles, $\theta_{c.m.} = 30^{\circ}, 90^{\circ}$ and $150^{\circ}$,
and two laboratory kinetic energies of the incoming nucleon, 
$E_{\mathrm{lab}} = 13~\mathrm{MeV}$ and $E_{\mathrm{lab}} = 135~\mathrm{MeV}$.
As in the two-body case, we use the chiral N$^4$LO 
and N$^4$LO$^+$ SMS potentials at $\Lambda = 450~\rm{MeV}$, and the OPE-Gaussian force.
The $d\sigma/d\Omega$ appears strongly correlated 
with $C_{xx}$ at $\theta_{c.m.} = 90^{\circ}$ and $\theta_{c.m.} = 150^{\circ}$ for $E_{\rm{lab}} = 13~\rm{MeV}$, 
and moderately correlated at the three scattering angles 
at $E_{\rm{lab}} = 135~\rm{MeV}$ for all the employed potentials. 
We find that for $\theta_{c.m.} = 30^{\circ}$ at the lower energy only a weak correlation for this pair of 3N observables exists.

Another picture emerges for the neutron analyzing power $A_{y}(n)$ and the deuteron vector analyzing power iT$_{11}$ 
which are strongly or moderately correlated, depending on the scattering angle and energy, 
see the 3rd and 4th rows of Fig.~\ref{fig4:scatt_3N}. 
The N$^{4}$LO (N$^{4}$LO$^{+}$) potential yields for $E_{\mathrm{lab}} = 13~\mathrm{MeV}$: at $\theta_{c.m.} = 30^{\circ}$ $r = 0.95$ ($r = 0.94$), 
at $\theta_{c.m.} = 90^{\circ}$ $r = 0.90$ $(r = 0.63)$, and at $\theta_{c.m.} = 150^{\circ}$ $r = 0.99$ $(r = 0.99)$. 
For $E_{\mathrm{lab}} = 135~\mathrm{MeV}$, the dependence between the observables looks very linear at $\theta_{c.m.} = 30^{\circ}$ 
and indeed the magnitudes of $r$ are: $\theta_{c.m.} = 30^{\circ}$ $r = 0.99$ $(r = 0.98)$ for the N$^{4}$LO (N$^{4}$LO$^{+}$) forces.
The same interactions lead at $\theta_{c.m.} = 90^{\circ}$ to $r = 0.60$ $(r = 0.33)$, 
and at $\theta_{c.m.} = 150^{\circ}$ $r = 0.71$ $(r = 0.77)$ with N$^{4}$LO (N$^{4}$LO$^{+}$). 

Finally, in the two bottom rows of Fig.~\ref{fig4:scatt_3N} we demonstrate yet another pattern for the correlation coefficient, which occurs for
the spin correlation coefficient $C_{XX}$ and the spin transfer coefficient $K^{y^{\prime}}_{y}(n)$.
Here, at lower energy and $\theta_{c.m.} = 30^{\circ}$ and $\theta_{c.m.} = 90^{\circ}$ the observables are strongly anticorrelated, while
at $\theta_{c.m.} = 150^{\circ}$ we observe strong positive correlation. This picture changes significantly when moving to the higher energy:
there is no correlation at $\theta_{c.m.} = 30^{\circ}$ but a strong positive correlation at $\theta_{c.m.} = 90^{\circ}$ and $\theta_{c.m.} = 150^{\circ}$.

Comparison between predictions of the chiral potentials and the OPE-Gaussian model reveals the very similar pattern of correlations.
The ``cloud'' of predictions has usually the same shape for all used forces what leads to comparable values of the correlation coefficient.
To give an example: in the case of $(A_{y}(n), \rm{iT}_{11})$ pair the OPE-Gaussian potential predicts the correlation coefficients:
$r = 0.89$ at $\theta_{c.m.} = 30^{\circ}$, $r = 0.95$ at $\theta_{c.m.} = 90^{\circ}$, 
and $r = 0.98$ at $\theta_{c.m.} = 150^{\circ}$ for $E_{\rm{lab}}=13~\rm{MeV}$ ($r = 0.99$ at $\theta_{c.m.} = 30^{\circ}$, $r = 0.61$ 
at $\theta_{c.m.} = 90^{\circ}$, and $r = 0.45$ at $\theta_{c.m.} = 150^{\circ}$ for $E_{\rm{lab}}=135~\rm{MeV}$). 

From other, not shown here, scatter plots it can be generally concluded that for the vast majority of 
selected angles at $E_{\rm{lab}} = 13~\rm{MeV}$ there is often a strong correlation, or less often observed 
moderate correlation. At the higher energy, $E_{\rm{lab}} = 135~\rm{MeV}$, we typically observe a weak correlation at 
almost all scattering angles, but there are exceptions that indicate a moderate correlation.
We can also conclude, that the correlation coefficients for Nd scattering observables predicted with the OPE-Gaussian, N$^4$LO, or N$^4$LO$^+$ forces
usually lead to the same qualitative conclusion about the correlation/uncorrelation. The biggest differences in predicted 
values of the correlation coefficients occur for uncorrelated or weakly correlated cases what is in agreement with our estimation
of correlation coefficients uncertainties. 

\begin{figure}[ht]
\centering
\includegraphics[scale=0.7,clip=true]{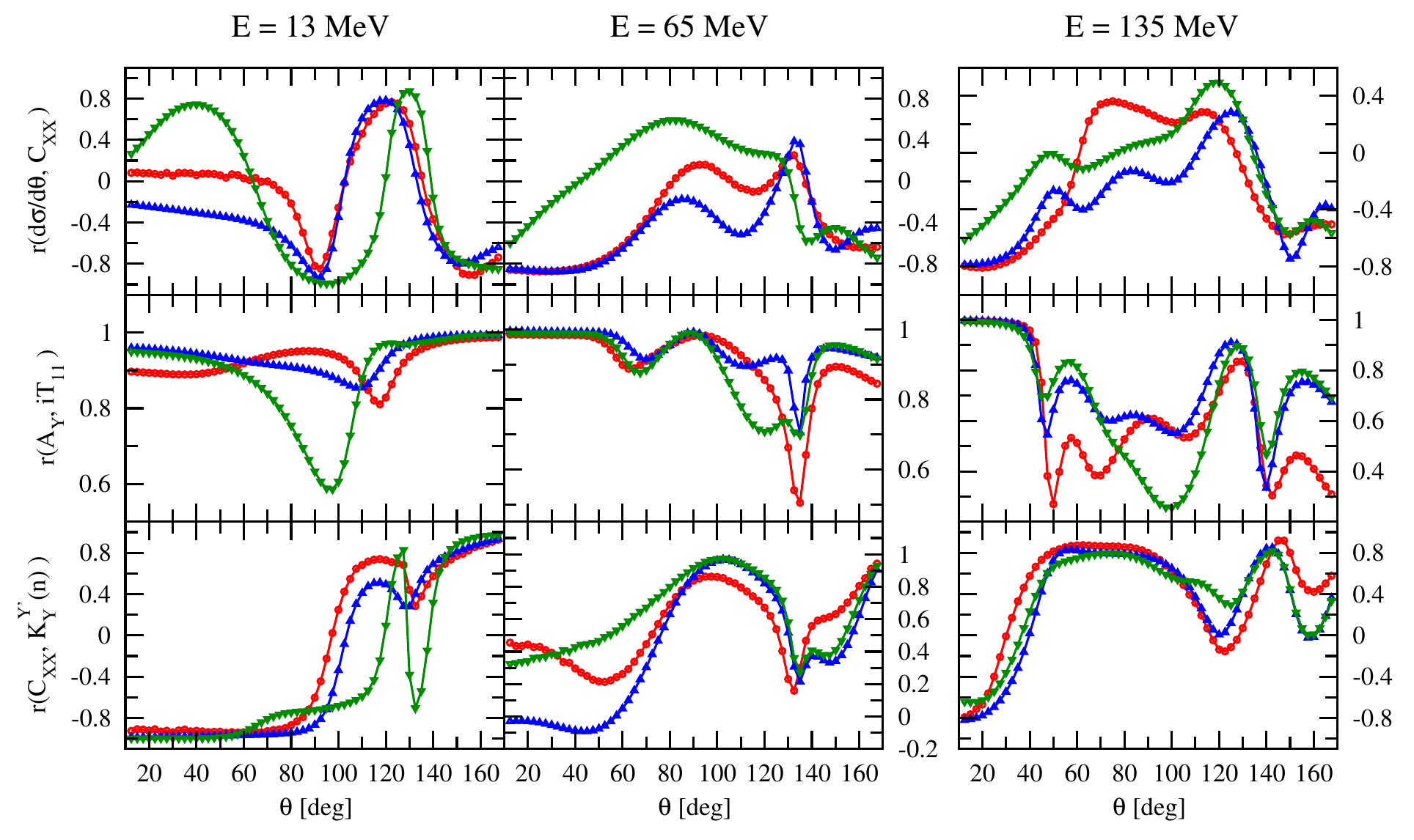}
\caption{
The angular dependence of correlation coefficients $r$ between $d\sigma/d\Omega$ and $C_{XX}$ (top),
the $A_Y(n)$ and $iT_{11}$ (the second row), and the $C_{XX}$ and $,K_Y^{Y'}(n)$ (bottom) in the $nd$ scattering
at the incoming neutron laboratory energy $E_{\rm{lab}} = 13~\rm{MeV}$ (left), $E_{\rm{lab}} = 65~\rm{MeV}$ (centre), and $E_{\rm{lab}} = 135~\rm{MeV}$ (right).
The red curve with circles, the blue curve with up-pointing triangles, and the green curve with down-pointing triangles represent predictions of the OPE-Gaussian,
the SMS N$^4$LO, and the SMS N$^4$LO$^+$ interactions, respectively.
}
\label{fig5:angular3N}
\end{figure}

Let us now turn to the angular dependence of the correlation coefficients.
We exemplify it in Fig.~\ref{fig5:angular3N},
for the $r(d\sigma/d\Omega, C_{xx})$, $r(A_Y(n), iT_{11})$ and $r(C_{XX},K_Y^{Y'}(n))$ at three scattering energies $E_{\rm{lab}} = 13, 65$ and 135 MeV. 
The predicted correlation coefficients are, as in the 2N case, complicated, energy-dependent, functions of the scattering angle
so again we restrict ourselves to qualitative conclusions only. 
The differential cross section $d\sigma/d\Omega$ is, in general, moderately correlated with $C_{xx}$ for the chiral SMS at both
presented  
orders, and for the OPE-Gaussian potential at all energies. However, we can also observe a weak or a strong correlation in
some ranges of the scattering angle. A weak correlation for this pair is observed at forward scattering 
angles at $E_{\rm{lab}} = 13~\rm{MeV}$, for $70^{\circ}<\theta_{c.m.}<150^{\circ}$ at $E_{\rm{lab}} = 65~\rm{MeV}$, 
and for $45^{\circ}<\theta_{c.m.}<140^{\circ}$ at $E_{\rm{lab}} = 135~\rm{MeV}$.
Strong anticorrelation is found at $E_{\rm{lab}} = 13~\rm{MeV}$ in the case of N$^{4}$LO$^{+}$ SMS predictions 
for $75^{\circ}<\theta_{c.m.}<115^{\circ}$ but with the increasing scattering 
angle it changes to strong correlation for $120^{\circ}<\theta_{c.m.}<140^{\circ}$. With increasing energy a strong correlation appears 
at forward/backward scattering angles. It is interesting that at the two lowest energies the N$^4$LO SMS and OPE-Gaussian predictions are close one to another
while the N$^{4}$LO$^+$ SMS results show a slightly different pattern. 
However, this is not very important for our qualitative study of the correlation coefficients
and does not affect our general conclusion about the weak/moderate dependence between these 3N observables.

The $(A_Y(n), iT_{11})$ pair remains strongly correlated at the two lower energies practically for all the scattering angles. 
At $E_{\rm{lab}} = 135~\rm{MeV}$
the correlation coefficient is close to one only for scattering angles below $\theta_{c.m.} \approx 40^{\circ}$ and 
correlation becomes moderate for bigger angles. Again at lower energies N$^4$LO$^+$ predictions are slightly different from
the N$^4$LO and OPE-Gaussian ones. 

Yet another pattern emerges for $r(C_{XX}, K_Y^{Y'}(n))$ and is displayed in the bottom row of Fig.~\ref{fig5:angular3N}. Here 
at the highest energy all predictions are close one to another and reveal weak, moderate or even strong correlation, depending on 
the range of the scattering angle. At two lower energies all predictions, in general, also show a similar behaviour; however
at $E_{\rm{lab}} = 13~\rm{MeV}$ the N$^4$LO$^+$ results show much stronger minimum around $\theta_{c.m.} = 135^{\circ}$ than the two
remaining predictions. 

Conclusions arising from this analysis of Fig.~\ref{fig5:angular3N} (and not shown here results for other pairs of 3N observables) 
are as follows:
\begin{enumerate}

\item the angular dependence of the correlation coefficients reveals complex structures for all pairs of observables 
and energies. These structures are a consequence of nonlinear dependence of the observables on potential parameters. 
In addition, the fact that the genuine potential parameters are fitted after the partial wave decomposition of the NN potential
makes this relation even more hidden. Also our method of calculating $r$, which is based on a sample of 50 predictions, 
introduces additional uncertainty. 
Nevertheless,
tests performed in the 2N system, and the fact that using  
different potentials (therefore different parameter spaces) leads to similar conclusions about the strength of correlations, 
seem to prove the correctness of the picture obtained.

\item in most cases, the N$^{4}$LO and N$^{4}$LO$^{+}$ yields similar correlation coefficients, although there are also pairs of 
3N observables for which they significantly differ for $120^{\circ}<\theta_{c.m.}<150^{\circ}$, especially 
at $E_{\mathrm{lab}} = 13~\mathrm{MeV}$. Beside $(C_{xx}, K^{y^{\prime}}_{y}(n))$ considered in Fig.~\ref{fig5:angular3N} these are: 
$(C_{zz}, K^{z^{\prime}}_{z}(d))$, $(K^{y^{\prime}}_{y}(n), K^{z^{\prime}}_{x}(n))$, $(K^{y^{\prime}}_{y}(n), K^{y^{\prime}}_{y}(d))$, 
$(K^{y^{\prime}}_{y}(n), K^{z^{\prime}}_{z}(d))$, $(K^{y^{\prime}}_{y}(n), K^{x^{\prime}y^{\prime}}_{x}(d))$, 
and $(K^{y^{\prime}}_{y}(n), K^{x^{\prime}y^{\prime}}_{z}(d))$.

\item at $E_{\mathrm{lab}} = 13~\mathrm{MeV}$ 3N observables are usually strongly correlated for $\theta_{c.m.} < 120^{\circ}$
when the chiral N$^{4}$LO$^{+}$ SMS potential is used. They become moderately or weakly correlated or even uncorrelated above 
this angle up to $\theta_{c.m.} \approx 150^{\circ}$ where again strong correlation is seen.

\item among the most easily experimentally available observables, i.e. the differential cross section, the neutron vector 
analyzing power $A_Y(n)$, the deuteron vector analyzing power $iT_{11}$ and the deuteron tensor analyzing powers T$_{20}$, T$_{21}$ and T$_{22}$, 
only the $(A_Y(n),iT_{11})$ pair shows strong correlation in the whole range of the scattering angle.
The (T$_{20}$, T$_{21}$) and (T$_{20}$, T$_{22}$) are strongly correlated for scattering angles $\theta_{c.m.}$ below below approx. $80^{\circ}$ at
the lowest studied energy $E_{\mathrm{lab}} = 13~\mathrm{MeV}$. At higher energies, all these observables, with the exception of  $(A_Y(n),iT_{11})$  
are characterized by weak correlation for a wide range of the scattering angle with correlation coefficients in the range of $-0.5 < r < 0.5$.
 
\item in the case of the N$^{2}$LO SMS force at all considered energies the correlation coefficient undergoes stronger 
changes with the scattering angle than the correlation coefficients computed with other potentials. The observed stabilization 
at higher orders shows that restricting calculations to the third order of the chiral expansion can be misleading for some 
correlation coefficients at the energies studied here.

\item the absolute value of the correlation coefficient decreases with increasing energy, which indicates weak correlation or even no correlation, 
with rare exceptions of $(C_{xx}, K^{y^{\prime}}_{y}(n))$ (Fig.~\ref{fig5:angular3N}) and $(C_{zz}, K^{y^{\prime}}_{y}(d))$ 
pairs. 
For other observables at $E_{\mathrm{lab}} = 65$ and 135 MeV strong/moderate correlation appears only in specific intervals 
of the scattering angles. 
This is true for all the used NN forces: the chiral SMS N$^{4}$LO, N$^{4}$LO$^+$, and the OPE-Gaussian potentials.
\end{enumerate}

\begin{figure}[ht]
%\captionsetup{width=\linewidth}
\centering
\includegraphics[scale=1.2,clip=true]{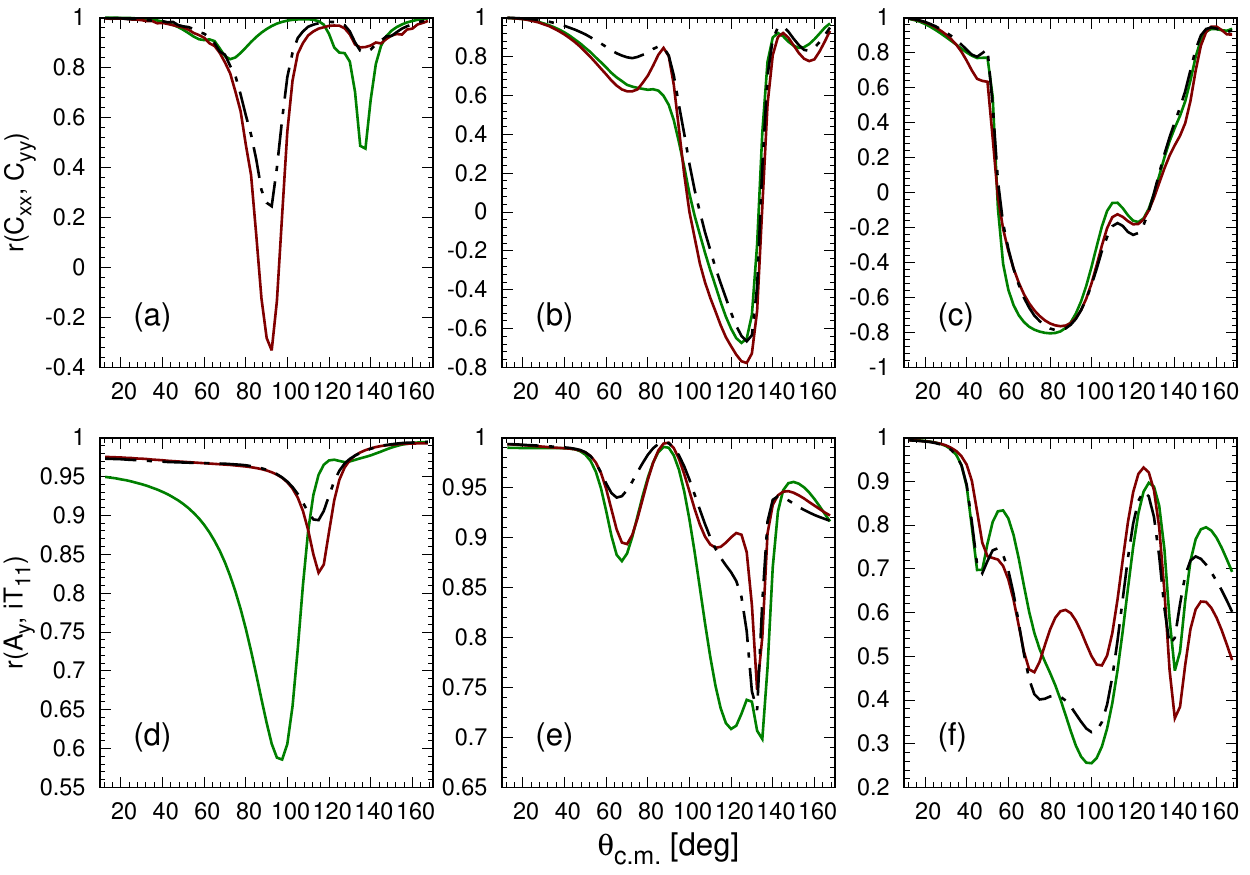}
\caption{The angular dependence of correlation coefficients $r(C_{xx}, C_{yy})$ (top) and $r(A_{y}(n), \mathrm{i}T_{11})$ (bottom) for different 
cutoff values (450 MeV --- green solid curve, 500 MeV --- maroon solid curve, and 550 MeV --- black dot-dashed curve) for the SMS force at 
N$^{4}$LO$^{+}$ order and at the incoming neutron laboratory energies $E_{\mathrm{lab}} = 13~\mathrm{MeV}$ (left), $E_{\mathrm{lab}} = 65~\mathrm{MeV}$ 
(center) and $E_{\mathrm{lab}} = 135~\mathrm{MeV}$ (right).}
\label{corr_Cxx_Cyy_Ay_iT11_N4LO+_L450_N4LO+_L500_N4LO+_L550} 
\end{figure}

All the SMS chiral force based results presented above 
have been obtained with the regularization parameter $\Lambda = 450~\mathrm{MeV}$. It is also interesting to study the 
sensitivity of correlation coefficients to the value of that cutoff parameter. 
Figure~\ref{corr_Cxx_Cyy_Ay_iT11_N4LO+_L450_N4LO+_L500_N4LO+_L550} demonstrates this dependence at N$^{4}$LO$^{+}$ using 
three values of $\Lambda$: 450, 500, and 550 MeV for the pairs $(C_{xx}, C_{yy})$ (the top row) and $(A_{y}(n), \mathrm{iT}_{11})$ (the bottom row). 
What is interesting - the cutoff dependence of $r(C_{xx}, C_{yy})$ is the most significant at the lowest energy $E_{\rm{lab}} = 13$~MeV,
while at the two higher energies predictions are close to each other, leading firmly to the same quantitative conclusion on 
correlation between these observables. In contrast, at $E_{\rm{lab}} = 13$~MeV and $\theta_{c.m.} \approx 90^{\circ}$ for $\Lambda=500$ and 550 MeV
we obtain a weak or moderate correlation, while for  $\Lambda=450$ the correlation remains strong. Also at $\theta_{c.m.} \approx 135^{\circ}$ 
the predictions based on $\Lambda=450$ cutoff yield moderate correlation, while the two other cutoff values suggest strong correlation.
Similar picture arises for $r(A_{y}(n), \mathrm{i}T_{11})$ where at the two higher energies, despite some differences between predictions, a qualitative
description of correlations is cutoff independent. At $E_{\rm{lab}} = 13$~MeV the SMS predictions with $\Lambda = 450~\mathrm{MeV}$ reveals
moderate correlation in comparison to strong correlation resulting when $\Lambda = 500~\mathrm{MeV}$ or $\Lambda = 550~\mathrm{MeV}$ is used.
At $E_{\mathrm{lab}} = 13~\mathrm{MeV}$ and for $\Lambda = 550~\mathrm{MeV}$ there is strong correlation in the whole 
range of $\theta_{c.m.}$ with the correlation coefficient dropping only to $\approx 0.9$ for $\theta_{c.m.} = 112.5^{\circ}$. 
The same is observed for $\Lambda = 500~\mathrm{MeV}$ 
but $r$ reaches a minimum of $\approx 0.83$ for $\theta_{c.m.} = 115^{\circ}$. At $E_{\mathrm{lab}} = 13~\mathrm{MeV}$ the 
cutoff $\Lambda = 450~\mathrm{MeV}$ provides a gradual decreasing of $r$ from 0.95 at $\theta_{c.m.} = 12.5^{\circ}$ to $r \approx 0.58$ 
at $\theta_{c.m.} = 97.5^{\circ}$, but starting from $\theta_{c.m.} \approx 100^{\circ}$ $r(A_{y}(n), iT_{11})$ increases. 

\begin{figure}[ht]
%\captionsetup{width=\linewidth}
\centering
\includegraphics[scale=1,clip=true]{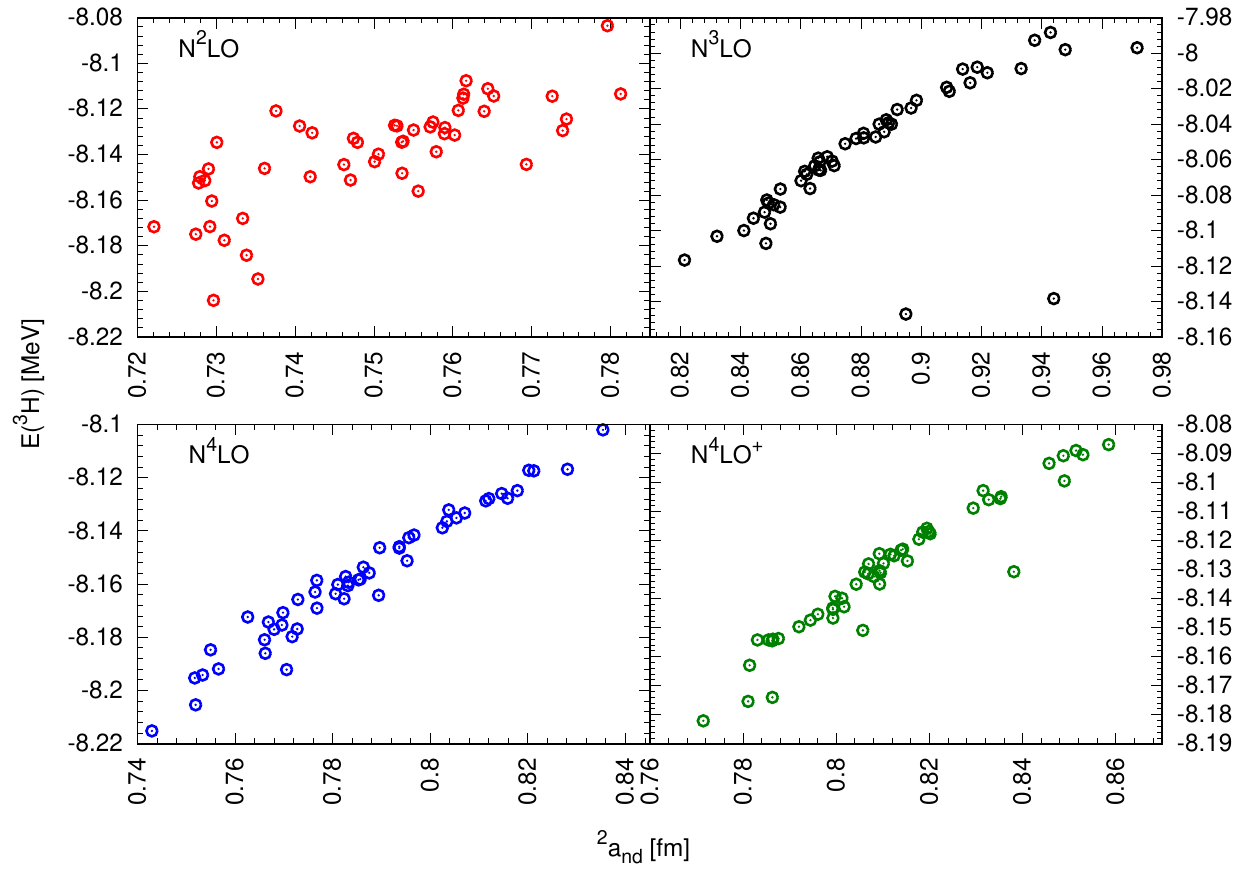}
\caption{Order-by-order results (N$^{2}$LO--N$^{4}$LO$^{+}$ with $\Lambda = 450~\mathrm{MeV}$) for the Phillips line E$(^3$H)-$^{2}$a$_{\rm{nd}}$.}
\label{phillips_line} 
\end{figure}

Another interesting case of correlation between observables is given by the so-called Phillips 
line~\cite{phillips1968consistency}, \cite{EFIMOV1985108} describing a dependence between the $^{3}$H binding energy 
and the nucleon-deuteron doublet scattering length, $^{2}$a$_{\rm{nd}}$. The Phillips line was observed in 
the past both in calculations with and without 3NF. As seen from Fig.~\ref{phillips_line} we reproduce the Phillips 
line using chiral N$^{3}$LO, N$^{4}$LO and N$^{4}$LO$^{+}$ SMS interactions with $\Lambda = 450~\mathrm{MeV}$. 
The correlation coefficient between these observables takes values of 0.75, 0.71 (0.97 after removing three outliers), 0.98, 
and 0.96 for predictions at N$^{2}$LO, N$^{3}$LO, N$^{4}$LO, and N$^{4}$LO$^{+}$, respectively. With the increasing chiral order, 
the values of these two observables change a little but are far from the experimental data 
($E(^{3}H) = -8.4820 \pm 0.0001~\rm{MeV}$~\cite{perez2014triton} and $^{2}$a$_{\rm{nd}} = 0.65\pm 0.04~\mathrm{fm}$~\cite{DILG1971208}). 
The observed discrepancy between our present predictions and the data is not surprising as it is well-known that both $^{3}$H binding energy and $^{2}$a$_{\rm{nd}}$ 
are strongly influenced by 3NF.

\begin{figure}[ht]
%\captionsetup{width=\linewidth}
\centering
\includegraphics[scale=1,clip=true]{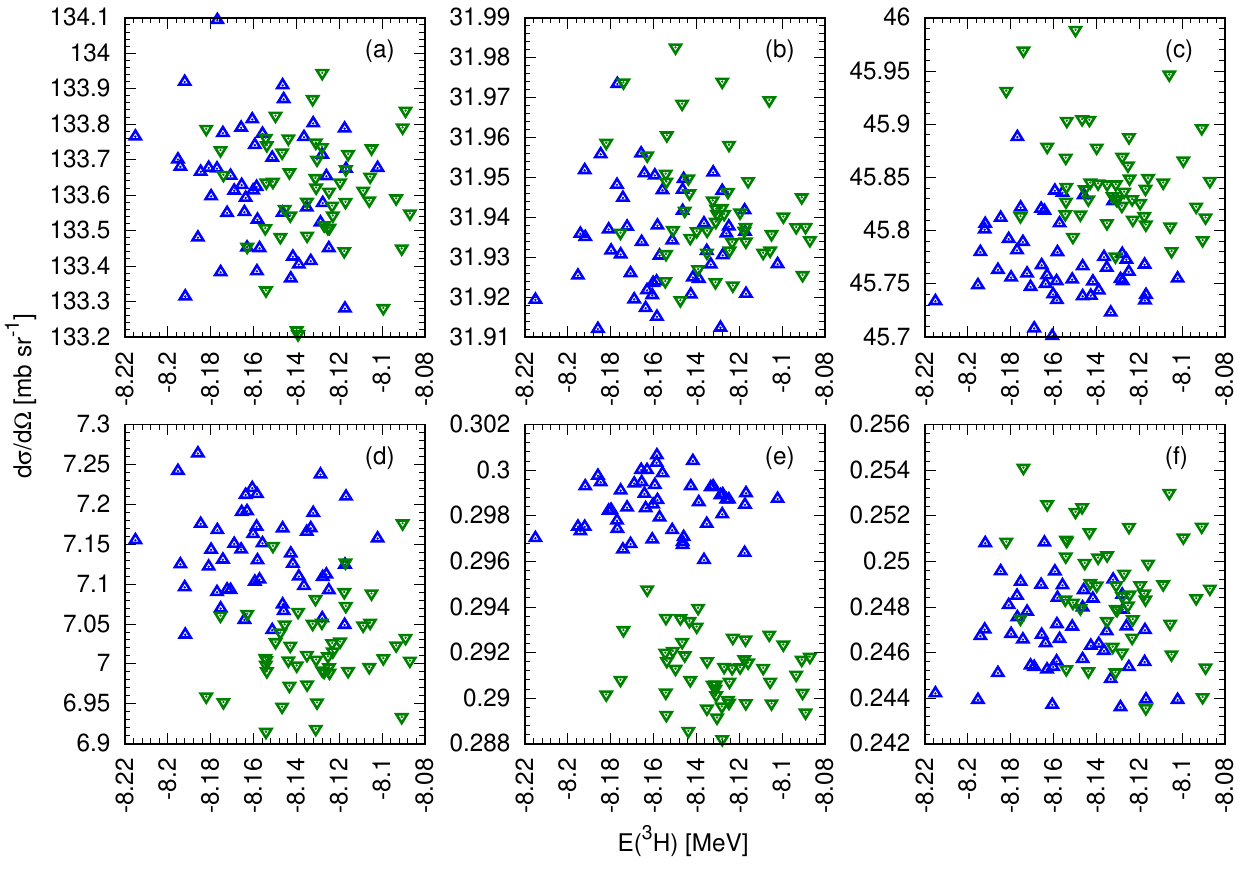}
\caption{A scatter plot between the $nd$ elastic differential cross section $d\sigma/d\Omega $ and the triton binding energy $E(^{3}H)$ at the c.m. scattering angle $\theta_{c.m.} = 30^{\circ}$ (left), $\theta_{c.m.} = 90^{\circ}$ (middle), and $\theta_{c.m.} = 150^{\circ}$ (right) and at the incoming neutron laboratory energy $E_{\rm{lab}} = 13~\rm{MeV}$ (top) and $E_{\rm{lab}} = 135~\rm{MeV}$ (bottom). 
The blue up-pointing triangles (the green down-pointing triangles) represent the chiral NN SMS predictions at N$^4$LO (N$^4$LO$^+$). } 
\label{corr_line_dsg_3h} 
\end{figure}

As mentioned in the introduction, the differential cross section at medium energy and the $^3$H binding energy $E(^3H)$ are nowadays 
used to fix parameters of the 3NF in the chiral SMS model. Now we are able to check if these two observables are uncorrelated. 
This is done in Fig.~\ref{corr_line_dsg_3h}. Indeed, the presented scatter plots confirm no correlation between these two observables. 
The correlation coefficient $r(d\sigma/d\Omega, E(^3H))$ is small at both energies and for all the scattering angles.
Its absolute value remains below $\left| r \right| = 0.28$. For not shown here energy 65 MeV $\vert r(d\sigma/d\Omega, E(^3H) \vert$ 
does not exceed $\left| r \right| \approx 0.3$. We expect that this picture will not change for complete predictions comprising 3NF.

\section{Summary and outlook}
\label{summary}

In the presented work we give a systematic analysis of the correlation coefficients among 2N and 3N observables.
We demonstrated that it is possible to analyze correlations among various 2N and 3N observables using the covariance matrices 
of potential parameters provided with the models of NN interactions from the Granada and Bochum-Bonn groups. 
Consequently, we showed that there are pairs of 2N spin observables for which an almost linear dependence exists, as documented by
correlation coefficients close to $\pm 1$.
For example, these are the cases of the $np$ depolarization R and the asymmetry $A^{\prime}$ or of the differential cross section
and the asymmetry $A^{\prime}$.
Correlations between given observables are observed both for the chiral SMS force (at N$^{3}$LO and beyond) and for the OPE-Gaussian potential. 
For some pairs of 2N observables, the angular dependence of the correlation coefficients depends strongly on 
the order of the chiral expansion as well as on the scattering energy. 

The same is true for elastic neutron-deuteron scattering. All correlation coefficients change in a complicated way with the 
scattering angle. For many pairs of 3N observables we found intervals of scattering 
angles where the correlation coefficient $\left|r\right| > 0.8$. We also studied the dependence of the correlation coefficients
on the order of the chiral expansion and on the scattering energy. For some pairs of the spin correlation and spin transfer 
coefficients, as well as for the analyzing powers, a moderate dependence on the chiral cutoff parameter $\Lambda$ is observed.
However, in most cases it does not change our qualitative conclusions on the correlation/uncorrelation 
between specific observables. Our results reflect a complex dependence of observables on potential parameters.
Our calculations reproduce the Phillips line E$(^3$H)-$^{2}$a$_{\rm{nd}}$ and support the practice of determining values of 
free parameters of the three-nucleon interaction from the uncorrelated differential cross section and the triton binding energy.

The next step in studies of correlations in few-nucleon systems is to investigate the sensitivity of a given observable
to the value of a specific potential parameter. In such a case one deals with nonlinear dependence of observables 
on many, mutually correlated, potential parameters. Thus much more sophisticated statistical tools than the sample correlation coefficient
must be applied. 
Information about sensitivity of some 2N or 3N observables to a given potential parameter can be used to improve a procedure of fixing potential parameters. 
Existence of strongly correlated observable-potential parameter pairs could also motivate experimental groups to perform precise 
measurements of such observable, especially if the experimental data are not yet available.

%\label{Summary}
% 

\acknowledgments
We would like to thank Dr E.Epelbaum and Dr P.Reinert from the R\"uhr-Universit\"at, Bochum, for providing us with the potential chiral subroutines.
This work was partly supported by the National Science Centre of Poland under grant number UMO-2020/37/B/ST2/01043 and by PL-GRID infrastructure.
The numerical calculations were partly performed on the supercomputer cluster of the JSC, J\"ulich, Germany.

%\bibliographystyle{unsrt}

%\bibliographystyle{prsty}
%\bibliographystyle{plain}
%\bibliographystyle{ieeetran}

%\bibliographystyle{apsrev4-1}

%\printbibliography
\bibliographystyle{apsrev4-1}
\bibliography{references}

\end{document}